\documentstyle[12pt,psfig,cite]{article}
\begin{document}
\renewcommand{\thefootnote}{\fnsymbol{footnote}}
\begin{titlepage}
%
%
\vspace*{10mm}
\Large
\begin{center}
Spontaneous Decoherence in Coupled Quantum Kicked Rotators 
\end{center}
\vspace{10mm}
\begin{center}
N.Tsuda {\small $^{\dag}$} \footnote[0]{E-mail address: ntsuda@theory.kek.jp}
and 
T.Yukawa {\small $^{\ddag , \dag}$} \footnote[0]{E-mail address: yukawa@theory.kek.jp}
\end{center}
\vspace{5mm}
\normalsize

\begin{center}
$^{\dag}$ Theory Division, Institute of Particle and Nuclear Studies, \\
KEK, High Energy Accelerator Research Organization\\
Tsukuba, Ibaraki 305 , Japan\\
\end{center}

\begin{center}
and
\end{center}

\begin{center}
$^{\ddag}$ Coordination Center for Research and Education,\\   
The Graduate University for Advanced Studies,\\
Hayama-cho, Miuragun, Kanagawa 240-01, Japan 
\end{center}
\vspace{1cm}
\parindent 5cm
\begin{abstract}
Quantum mechanical behavior of coupled $N$-kicked rotators is studied. 
In the large $N$ limit each rotator evolves under influence of the mean-field 
generated by surrounding rotators. 
It is found that the system spontaneously generates classical chaos in the 
large $N$ limit when the system parameter exceeds a critical value.
Numerical simulation of a quantum rotator coupled to a classical rotator
supports this idea. 
\end{abstract}
\end{titlepage}
 
\section{Introduction}
In spite of numerous efforts to define quantum chaos, there seems to 
exist many counter examples of genuine chaos in quantum systems.
Among those the quasi-periodic time evolution of isolated bound systems 
and the limited diffusion due to the momentum localization in the 
periodically kicked quantum rotator are typical examples \cite{CH0,CH01,CH02}.  
Although they are rather persuasive to give up searching quantum chaos, 
we still reserve a good reason to believe the seeds of chaos should exist 
in quantum system which grow to classical chaos.
One way to observe it is in the time evolution of the Wigner function;
\begin{equation}
\frac{\partial f}{\partial t}=L[f].
\end{equation}
When the quantum Liouville operator $L$ is expanded in terms of the 
Planck constant as
\begin{equation}
L=L_{0}+\hbar^{2}L_{1}+\cdots ,
\end{equation}
the lowest order term  generates the classical evolution,
\begin{equation}
L_{0}[f] = \sum_{i} \{H,f\}^{(i)},
\end{equation}
where 
$\{H,f\}^{(i)}$ represents the Poisson bracket with respect 
to the $i$-th canonical variables ($q_{i},p_{i}$) of a system with 
the Hamiltonian $H(\{q_{i}\},\{p_{i}\})$. 
Therefore, it is natural to search the origin of chaos in the region of 
$\hbar \sim 0$, $i.e.$ the semi-classical limit.
In fact, recent intensive investigations \cite{CH1,CH11} have 
revealed rich structures in this limit.  

There is another way to observe the emergence of chaos.
If it exists in the system which involves many degrees of freedom.
Since the most characteristic property of a macroscopic system is the 
large number of degrees of freedom involved in describing the motion, 
it is natural to consider a system with infinite number of degrees of 
freedom as the realistic classical limit. In this case we expect that 
quantum mechanical phases of each degree of freedom interfere mutually, 
and are wiped out in chaotic environment resulting the classical behavior 
of the Wigner function, {\it i.e.} the classical distribution function
\cite{DSC}.

When the number of degrees of freedom $N$ involved in a system tends 
to infinity, the mean-field approximation often becomes exact due to 
the large number theorem. In such a case the Wigner function can be
approximated in a product form\cite{TY},
\begin{equation}
f(\{q_{i}\},\{p_{i}\};t) \approx \prod^{N}_{i=1} f_{i}(q_{i},p_{i};t),
\end{equation}
of functions corresponding to the $N$ sub-systems. 
The Wigner function for the $i$-th sub-system then evolves as
\begin{equation}
\label{eq:Liouville_eq}
\frac{\partial f_{i}}{\partial t} = L_{i}^{(t)}[f_{i}],
\end{equation}
with the appropriately defined mean-field operator, $L^{(t)}_{i}$. 
In general, this operator depends on the temporal distributions of 
surrounding sub-systems. Thus the mean-field Liouville equation 
(\ref{eq:Liouville_eq}) becomes non-linear and time-dependent. 
It is our expectation that the non-linearity and the time-dependence induce 
chaos, and chaos destroys quantum phases.      

In the next section we shall describe time evolution of the Wigner 
function for a system of coupled kicked rotators, and the mean-field 
approximation is introduced. 
In the third section the mean-field equation is replaced by an equivalent 
system of a quantum kicked rotator coupled with a classical kicked rotator 
representing the environment.
And numerical simulation of this system is examined. 
The idea of spontaneous decoherence is discussed based on the numerical 
calculation in the last section.
%

\section{Large $N$ limit and the statistical mean-field approximation}
Let us examine the scenario more explicitly by taking the 
coupled $N$-kicked rotator model \cite{CH2} as an example. 
 The Hamiltonian is assumed to be
\begin{equation}
\label{eq:hamiltonian}
H = \sum^{N}_{i=1}\{\frac{1}{2} t_{ii} p_{i}^{2} + \Delta(t)\lambda_{i}
\cos \theta_{i} \}+\frac{1}{2} \sum^{N}_{i \neq j} t_{ij} p_{i} p_{j},
\end{equation}
where
$$
\Delta(t) = \sum^{\infty}_{n=-\infty} \delta(t-n)
$$
is the periodical kick. In this model couplings of $N$ kicked rotators 
enter through off-diagonal kinetic terms. Couplings through the kick 
term \cite{CH3} may be converted to this form through a linear 
transformation of angle variables. 
The classical equation of motion is known as the {\it standard map} 
which is the mapping between canonical coordinates just after
kicks at times $t$ and $t+1$,
\begin{equation}
\left\{
\begin{array}{ll}
\theta^{(t+1)}_{i}\:\: = & \theta^{(t)}_{i} + \sum_{j} t_{ij} p^{(t)}_{j} \\
p^{(t+1)}_{i}\:\: = & p^{(t)}_{i} + \lambda_{i} \sin \theta^{(t+1)}_{i}  .
\end{array}
\right. \;\;\;\;\;\; (t:integer)
\label{eq:standard_map}
\end{equation}
We shall choose the magnitudes of matrix elements of the kinetic term as
\begin{equation}
\begin{array}{ll}
t_{ii} \sim {\cal O} (1) , \\ 
t_{ij} \sim {\cal O} (\frac{1}{N^{\nu}}) ~\mbox{for i $\neq$ j},
\end{array}
\end{equation}
where the exponent $\nu$ will be determined later 
so that the system is non-trivial and converging in the large $N$ limit.

Evolution of the Wigner function is obtained by solving the Schr\"odinger 
equation. 
The relation between Wigner functions just after kicks at times $t$ and 
$t+1$ is called the quantum mapping, and it can be written as
\begin{equation}
f^{(t+1)}(\{\theta_{i}\},\{p_{i}\}) =  UVf^{(t)}
(\{\theta_{i}\},\{p_{i}\})
\label{eq:quantum_map}
\end{equation}
with the mapping operators $U$ and $V$ defined by
\begin{equation}
U = \exp(-\sum_{i} \lambda_{i} \sin \theta_{i} 
\frac{D}{D p_{i}}) ,
\label{eq:Up_ope}
\end{equation}
and
\begin{equation}
V = \exp(-\sum_{ij} t_{ij} p_{i} \frac{\partial}
{\partial \theta_{j}}) .
\label{eq:Ut_ope}
\end{equation}
Here, the operator $\frac{D}{D p}$ appeared in  $U$,
\begin{equation}
\frac{D}{D p} \equiv \frac{1}{\hbar}
(e^{ \frac{\hbar}{2} \frac{\partial}{\partial p}}-
 e^{-\frac{\hbar}{2} \frac{\partial}{\partial p}}) ,
\end{equation}
is a difference operator, and quantum effects in the 
evolution equation enter only through 
this operator. Since it approaches to the differential operator 
in the classical limit as
\begin{eqnarray}
\frac{D}{D p}f(p) & = & \frac{1}{\hbar}
[f(p + \frac{\hbar}{2})-f(p - \frac{\hbar}{2})]  \nonumber \\
& \sim & \frac{\partial}{\partial p}f(p) , (\hbar \rightarrow 0)
\label{eq:diff_op}
\end{eqnarray}
the quantum mapping reduces to the well-known standard map of classical 
distribution function,
\begin{equation}
f^{(t+1)}(\{\theta_{i}\},\{p_{i}\}) \approx f^{(t)}(\{\theta_{i}-\sum_{j}
t_{ij}(p_{j}-\lambda_{j} \sin \theta_{j})\},\{p_{i}-\lambda_{i} \sin 
\theta_{i}\})
\end{equation}
in the $\hbar \to 0$ limit.

Although the difference between classical and quantum evolutions looks 
very small from eqs.(\ref{eq:Up_ope}) and(\ref{eq:diff_op}) at a glance, 
the consequence of actions of the operator $U$ on the exponential function 
may be instructive.
In quantum mechanical case it acts as
\begin{equation}
e^{(-a \frac{D}{D p})} e^{ikp} = e^{ik\{p-a \varphi(k)\}} ,
\end{equation}
with   
\begin{equation}
\varphi(k) = \frac{2}{\hbar k}\sin \frac{\hbar k}{2} ,
\end{equation}
while in the classical limit it reduces to the standard shift operator
along the momentum axis,
\begin{equation}
e^{-a\frac{\partial}{\partial p}} e^{ikp} = e^{ik(p-a)} ,
\end{equation}
{\it i.e.} the operator $\frac{D}{D p}$ behaves as the usual 
shift operator only for small wave number components. 
For large wave numbers shifts are suppressed by the factor
$1/k$ and direction of the shift alternates. 
This implies that rapid variations of the Wigner function along the 
p-axis are suppressed, and consequently chaotic folding of the 
distribution function terminates at fine scale of the order $\hbar$. 

Now, let us consider the large $N$ limit where we expect that the 
mean-field approximation works well. 
Let us suppose at a certain time $t$ the Wigner function is written in a 
product form,
\begin{equation}
f^{(t)}(\{\theta_{i}\},\{p_{i}\}) = \prod^{N}_{i=1} f^{(t)}_{i}
(\theta_{i},p_{i}).
\label{eq:mfwf} 
\end{equation}
Integrating the evolution equation(\ref{eq:quantum_map}) over all phase 
volume except the $i$-th canonical coordinates, we obtain the equation 
for the reduced Wigner function of the $i$-th sub-system:
\begin{eqnarray}
\label{eq:red_Liouville}
\lefteqn{f_{i}^{(t+1)}(\theta_i,p_i) = 
 \exp(- \lambda_{i} \sin \theta_{i} \frac{D}{D p_{i}})
\exp(-t_{ii} p_{i} \frac{\partial}{\partial \theta_{i}})}\nonumber \\
\hspace*{-2cm}& &\times \int \prod_{j(\neq i)} d\theta_j dp_j 
\exp(-\sum_{j (\neq i)} t_{ij} p_{j} 
\frac{\partial}{\partial \theta_{i}}) f^{(t)}(\{\theta_{j}\},\{p_{j}\}) ,
\end{eqnarray}
where the reduced Wigner function has been defined at $t+1$ as
\begin{equation}
f^{(t+1)}_{i}(\theta_{i},p_{i}) = \int \prod_{j(\neq i)} d\theta_j dp_j 
 f^{(t+1)}(\{\theta_{j}\},\{p_{j}\}) , 
\end{equation}
which will turn out to be the solution of the equation which we may call 
the {\it statistical} mean-field equation evolved from 
$f^{(t)}_{i}(\theta_{i},p_{i})$ of eq.(\ref{eq:mfwf}).

Expanding the exponential function in the integrand of the r.h.s. of 
eq.(\ref{eq:red_Liouville}) we obtain
\begin{eqnarray}
\lefteqn{\left[ 1-\sum_{j(\neq i)} t_{ij} \langle p_j \rangle _j^{(t)} 
\frac{\partial}{\partial \theta_{i}} \right. } \nonumber \\
&+& \left. {1 \over 2} \sum_{j,k(j\neq k,\neq i)} t_{ij} t_{ik} 
\langle p_j \rangle _j^{(t)} \langle p_k \rangle _k^{(t)}
\frac{\partial ^2}{\partial \theta_{i}^2} + {1 \over 2} \sum_{j(\neq i)} 
t_{ij}^2  \langle p_j^2 \rangle _j^{(t)} \frac{\partial ^2}
{\partial \theta_{i}^2} \right] f_i^{(t)}(\theta_i,p_i),\nonumber
\end{eqnarray}
for the first and second order terms, where we have written
\begin{equation}
\langle p_{j}^{n} \rangle^{(t)}_j = \int d\theta_j dp_j p_{j}^{n} 
f^{(t)}_j(\theta_j,p_j) .
\end{equation}
Writing back this expression to the exponential form we have
$$
\exp(-\sum_{j(\neq i)} t_{ij} \langle p_j\rangle _j^{(t)}
\frac{\partial}{\partial \theta_{i}}
+ \sum_{j(\neq i)} t_{ij}^2 \sigma_j^{(t)}
\frac{\partial ^2}{\partial \theta_{i}^2})f_i^{(t)},
$$
for the first two terms of the cumulant expansion.
Here, $\sigma_j^{(t)}$ is the momentum variance of the $j$-th sub-system,
\begin{equation}
\sigma_j^{(t)} = \langle p_j^2 \rangle _j^{(t)} 
- \{\langle p_j \rangle _j^{(t)} \}^2.
\end{equation}
We have chosen that the coupling strength $t_{ij} (i \neq j)$ to be  
${\cal O} (N^{-\nu})$, and thus the $n$-th cumulant ({\it i.e.} the term 
involving $\frac{\partial ^n}{\partial \theta_{i}^n}$) is of the order of 
$N^{1-n\nu}$. 
The parameter $\nu$ classifies nature of the system into the following 
cases in the $N \to \infty$ limit:
\begin{itemize}
\item [i)] $\nu > 1$, any cumulant converges to $0$ and the system 
becomes a set of independent kicked rotators.
\item [ii)] $\nu = 1$, only the first cumulant 
$\sum_{j(\ne i)}t_{ij} \langle p_{j} \rangle_{j}^{(t)}$ survives and the 
{\it ordinary} mean-field Hamiltonian, 
\begin{equation}
\bar{h}_{i} = h_{i} + \sum_{i \ne j} p_{i} t_{ij} 
\langle p_{j} \rangle_{j}^{(t)},
\end{equation}
drive the reduced Wigner function $f_{i}^{(t)}$.
\item [iii)] $\frac{1}{2} < \nu < 1$, the ordinary mean-field theory is valid 
assuming $\sum_{j(\ne i)}t_{ij} \langle p_{j} \rangle_{j}^{(t)}$ remains finite.
By transforming the system in a rotating frame with an appropriate momentum 
$\bar{p}$, the fluctuation 
$\sum_{j(\ne i)} t_{ij} (\langle p_{j} \rangle_{j}^{(t)} - \bar{p})$ can be kept finite.
\item [iv)] $\nu = \frac{1}{2}$, the statistical mean-field theory is 
valid providing 
$\sum_{j(\ne i)} t_{ij} (\langle p_{j} \rangle_{j}^{(t)} - \bar{p})$
kept finite.
\item [v)] $\nu < \frac{1}{2}$, the second cumulant diverges and the 
system becomes unphysical.
In this case it does not make any sense to consider the reduced Wigner 
function.
\end{itemize}

The statistical mean-field equation is given by
\begin{eqnarray}
f_{i}^{(t+1)}(\theta_i,p_i) 
= \exp(- \lambda_{i} \sin \theta_{i} \frac{D}{D p_{i}})
\exp(-t_{ii} p_{i} \frac{\partial}{\partial \theta_{i}})
\nonumber \\
\exp(-\sum_{j (\neq i)} t_{ij} \langle p_{j}\rangle _j^{(t)}
 \frac{\partial}{\partial \theta_{i}}
 +\sum_{j(\neq i)} t_{ij}^2 \sigma_j^{(t)}
\frac{\partial ^2}{\partial \theta_{i}^2}) 
f_{i}^{(t)}(\theta_i,p_i).
\end{eqnarray}
Since evolution operator of the $i$-th subsystem involves only the $i$-th 
canonical coordinates product form of the Wigner function will remains at 
$t+1$:
\begin{equation}
f^{(t+1)}(\{\theta_{i}\},\{p_{i}\}) = \prod^{N}_{i=1} f^{(t+1)}_{i}
(\theta_{i},p_{i}).
\label{eq:mfwf2} 
\end{equation}
The proof that the product eq.(\ref{eq:mfwf2}) is indeed the solution of 
the evolution equation (\ref{eq:quantum_map}) is given in ref.\cite{TY}.

For studying asymptotic behavior of the momentum distribution it is more 
convenient to consider the Fourier transform of the reduced Wigner function 
in the statistical mean-field approximation,
$$
\tilde{f}^{(t)}_{i}(m_{i},k_{i}), 
$$
where $m_{i}$ and $k_{i}$ are the Fourier parameters corresponding to the 
variables $\theta_{i}$ and $p_{i}$, respectively. 
The momentum distribution function for the $i$-th sub-system is then 
readily expressed as
\begin{equation}
f^{(t)}_{i}(p) = \int^{2\pi}_{0} \frac{dk}{2 \pi \hbar} 
e^{-ikp/\hbar} \tilde{f}^{(t)}_{i}(0,k).
\label{eq:mom_dist}
\end{equation}
The evolution equation in this representation is written as
\begin{equation}
\tilde{f}^{(t+1)}_{i}(m_{i},k_{i}) = 
\sum_{m_{i}'} J_{m_{i}-m_{i}'}
(\frac{2 \lambda_{i}}{\hbar} \sin \frac{\hbar  k_{i}}{2}) 
e^{-im_{i}'g^{(t)}_{i}-{1 \over 2} m_i'^2 \Sigma^{(t)}_i} 
\tilde{f}^{(t)}_{i}(m_{i}',k_{i}-m_{i}'),
\label{eq:evolution_sub}
\end{equation}
where $J_{m}(k)$ is the Bessel function of the order m, 
and we have defined
\begin{equation}
g^{(t)}_{i}=\sum_{j (\neq i)} t_{ij} \langle p_{j}\rangle _j^{(t)},
\label{eq:mf_strength}
\end{equation}
\begin{equation}
\Sigma^{(t)}_{i}=\sum_{j(\neq i)} t_{ij}^2 \sigma_j^{(t)}.
\end{equation}

For the asymptotic behavior of $\Sigma^{(t)}_i$, there will exist at least 
two cases depending on the nature of environment:
\begin{eqnarray}
\sigma^{(t)}_i &\rightarrow& const. ({\rm regular}) \nonumber\\
               &\rightarrow& d_i t .~~~ ({\rm chaotic})
\label{eq:variant}
\end{eqnarray}
When the environment is chaotic the sum over $m_i$ in 
eq.(\ref{eq:evolution_sub}) will be dominated asymptotically by the $m_i = 0$ 
components.
The asymptotic form of the evolution equation then becomes
\begin{equation}
\tilde{f}^{(t+1)}_{i}(m_{i},k_{i}) \sim
 J_{m_{i}}
(\frac{2 \lambda_{i}}{\hbar} \sin \frac{\hbar  k_{i}}{2})  
\tilde{f}^{(t)}_{i}(0,k_{i}) ,
\end{equation}
whose solution is now trivial. 
Especially the $m_i = 0$ components are given as
\begin{equation}
\tilde{f}^{(t)}_{i}(0,k_{i}) \sim
 \{ J_{0}
(\frac{2 \lambda_{i}}{\hbar} \sin \frac{\hbar  k_{i}}{2}) \}^{t}  
\tilde{f}^{(0)}_{i}(0,k_{i}) .
\end{equation}
Inserting this solution into eq.(\ref{eq:mom_dist}) together with 
help of the cental limit theorem we obtain the asymptotic form of 
the momentum distribution function as 
\begin{equation}
f^{(t)}_i(p) \sim [{1 \over \pi t \lambda_{i}^{2}}]^{{1 \over 2}}
e^{-{p^2 \over t \lambda_{i}^2}} ,
\label{eq:cmom_dist}
\end{equation}
which is the same function as the classical distribution 
of the chaotic limit.
The $m = 0$ dominance suggests that when the environment 
become chaotic the system starts behaving classically
due to the phase decoherence.

\section{An equivalent model and numerical results}
We have seen that the $N$-coupled kicked rotator will fall into chaos in 
the large $N$ limit in cases when the statistical mean-field approximation 
works well.
Under those circumstances it becomes possible to study, instead of treating 
whole system at once, just by picking up any one of the rotators as an 
object system which we simply call the system($S$) under influence of other 
rotators as the environment($E$).
Interaction between $S$ and $E$ enters as an external field with strength 
proportional to the state averaged momentum in our model.
Although the mean-field approximation makes it possible to perform the 
large $N$ calculation as many as $N\approx 100-1000$ by using large scale 
parallel processors, we leave such huge simulations as the future project. 
Instead, we assume the mean-field strength (\ref{eq:mf_strength}) is 
replaced by a classical system with appropriate initial ensemble. 
We choose the ensemble so that it become equivalent to the statistical 
mean-field theory.

Now, let us consider a quantum kicked rotator($S$) with the Hamiltonian, 
\begin{equation}
H_S = \frac{1}{2} p^2 + \Delta(t) \lambda_S 
\cos \theta +  g^{(t)} p
\label{eq:qc_hamiltonian}
\end{equation}
under the influence of classical environment($E$) through 
 $g^{(t)}$.
 This system evolves as
\begin{equation}
f_S^{(t+1)}(\theta,p) = \exp(-\lambda_S \sin \theta \frac{D}{D p}) 
\exp(- p \frac{\partial}{\partial \theta})\\
\exp(- g^{(t)} \frac{\partial}{\partial \theta})f_S^{(t)}(\theta,p) ,
\label{eq:qc_evolution}
\end{equation}
or by the Fourier transform it reads
\begin{equation}
\tilde{f}^{(t+1)}_{S}(m,k) = 
\sum_{m'} J_{m-m'}
(\frac{2 \lambda_S}{\hbar} \sin \frac{\hbar k}{2}) 
e^{-im'g^{(t)}} 
\tilde{f}_{S}^{(t)}(m',k-m') .
\end{equation}
After taking average over the initial ensemble the evolution equation 
is written as
\begin{equation}
E[\tilde{f}_S^{(t+1)}(m,k)] = \sum_{m'} J_{m-m'}
(\frac{2 \lambda_S}{\hbar} \sin \frac{\hbar k}{2}) E[e^{-im'
g^{(t)}} \tilde{f}_S^{(t)}(m',k-m')],
\end{equation}
where we write the ensemble averaged quantities by $E[\; \;]$. 
We assume the ensemble of mean-field couplings $\{g^{(t)}\}$ 
 to be  Gaussian with the correlation,
\begin{equation}
E[g^{(t)} g^{(t')}] - E[g^{(t)}] E[g^{(t')}]= \delta_{tt'} \sigma^{(t)} .
\label{eq:correlation}
\end{equation}
In this case the correlation between $g^{(t)}$ and 
$\tilde{f}^{(t)}$ vanishes, and we have 
\begin{equation}
E[\tilde{f}^{(t+1)}_{S}(m,k)] = \\
\sum_{m'} J_{m-m'}
(\frac{2 \lambda_S}{\hbar} \sin \frac{\hbar}{2} k) 
e^{-im'E[g^{(t)}]-{1 \over 2} m'^2 \sigma^{(t)}} 
E[\tilde{f}_S^{(t)}(m',k-m')] .
\label{eq:e_tilde}
\end{equation}
This equation is equivalent to the statistical mean-field equation
(\ref{eq:evolution_sub}).

Our next task is to give $g^{(t)}$ explicitly in order to 
solve the evolution eq.(\ref{eq:qc_evolution}). One of the possible
 choice of $g^{(t)}$
 which possesses both properties eq.(\ref{eq:variant}) and
(\ref{eq:correlation}) is the momentum generated by 
the classical kicked rotator,
\begin{equation}
\left\{
\begin{array}{ll}
\phi^{(t+1)}\:\: = & \phi^{(t)} + g^{(t)} \\
g^{(t+1)}\:\: = & g^{(t)} + \lambda_E \sin \phi^{(t+1)}  
\end{array}
\right .
\label{eq:sdmap}
\end{equation}
with a set of initial values $(\phi^{(0)},g^{(0)})$ as the ensemble.
Iterating eq.(\ref{eq:qc_evolution}) together with eq.(\ref{eq:sdmap})
the behavior of $S$ will show the quantum-classical
transition depending on the environment.

Before examining results of the numerical calculation let us remark
the consistency of the choice of $g^{(t)}$. 
The momentum distribution of $E$ in the mean-field approximation 
is given by
\begin{equation}
f^{(t)}_E(p_E)=\int \prod_{j (\neq S)}dp_j \delta (p_E-\sum_{j (\neq S)}p_j)
\prod_{j (\neq S)} f^{(t)}_j(p_j) .
\end{equation}
If each rotator is in a chaotic state its momentum distribution is known
to be
\begin{equation}
f^{(t)}_j(p) \propto e^{-{p^2 \over td_j}} 
\end{equation}
asymptotically. 
In this case  $f^{(t)}_E(p_E)$ also
has the same form with the diffusion constant being 
\begin{equation}
d_E=\sum_{j (\neq S)}t_{Sj}^2 d_j,
\end{equation}
which is ${\cal O}(1)$ for our choice of $t_{Sj} \approx {\cal O}
({1 \over N^{1/2}})$.

We shall now examine numerical results which support 
the scenario described in the last section. 
We solve the Schr\" odinger equation with the Hamiltonian
(\ref{eq:qc_hamiltonian}) with $g^{(t)}$ generated from
the standard map (\ref{eq:sdmap}). At each time step we measure the  
momentum fluctuation of $S$,
$$  
E[\langle p^{2} \rangle^{(t)}],
$$
averaged over the initial ensemble which consists
typically 20 members for each set of parameters $\lambda_S,\lambda_E$ 
and $\mu$.
As in the classical rotator chaos is measured by the diffusion constant 
defined by the rate of momentum fluctuation.

In order to see the effect of chaotic environment the system is maintained 
in chaos by choosing $\lambda_S=6.0$, and varying the environment 
$\lambda_E$ (Fig.1).
The coupling strength is chosen to be $\mu = 0.01$.
As we have expected the energy diffusion starts precisely when the 
environment become chaotic, $i.e.$ $\lambda_E \approx 1.0$ .
To make sure that the increase of kinetic energy is not due to the direct 
penetration of energy from $E$ via coupling, we made the similar 
calculation by keeping the system in the regular regime, $\lambda_S = 0.8$.
No energy diffusion has been observed in this case.

Choice of the coupling strength $\mu=0.01$ comes from the requirement that 
the property of $S$ should not be altered significantly by the environment.
Weakness of the coupling effect is clear if we compare diffusion constants 
of an isolated classical rotator which is $18.0$ for $\lambda = 6.0$.
No significant difference is observed when the coupling strength is less 
than $0.01$.
Although the effect of environment is small on the classical system, it has 
the essential action on the quantum system. 
It is seen in Fig.2 where $S$ and $E$ are kept in chaos 
($\lambda_S = \lambda_E = 6.0$) and the coupling strength is varied.
Above the lower critical coupling $\mu \approx 0.001$ the environment 
effect appears in the diffusion constant of $S$.

Next, we observe the transition of $S$ from the regular motion to chaos under 
the influence of the chaotic environment with keeping $\lambda_E=6.0$ (Fig.3). 
At the classical critical strength $\lambda = 1.0$ the quantum system persists 
to be diffusion-less.
It is only after $\lambda \approx 3.0$ the system shows diffusion.
Change of the evolution from quantum to classical by the effect of $E$ 
can also be seen in the momentum distribution. 
Figs.4 (a) and (b) correspond to the regular motion and the chaotic 
regime, respectively. In the regular case the momentum localization 
persists during the simulation time up to $t<300$, while in the chaotic 
case the momentum distribution changes following the classical diffusion 
law described by eq.(\ref{eq:cmom_dist}).

\section{Spontaneous decoherence -A conclusion-}
It is well-known that a single quantum kicked rotator shows only limited 
energy diffusion. The limited diffusion is considered to be generic in 
quantized systems with a few degrees of freedom whose motion show chaos in 
the classical evolution. We have searched the source of classical diffusion 
in systems with infinite number of degrees of freedom. 
These system can be treated in the statistical mean-field approximation, 
where we can pick any one of the sub-system as an object system and the 
rest as the environment. 
Energy diffusion has been observed in the quantum system when the 
environment is in classical chaos. 
Momentum distribution of the quantum system is also same as one which we 
normally observe in a classical chaotic system. 
Since each building block of the environment is the same type of quantum 
system, the transition from the quantum behavior to the classical behavior 
seen in the diffusion should occur spontaneously.

One may wonder then where is the seed of chaos in linear dynamical system 
like quantum mechanical system. It is certainly very difficult to point out 
the cause of chaos unlike classical dynamical system where the existence of 
homoclinic point means chaos. 
Although we have no rigorous proof for the existence of chaos in quantum 
systems in the limit of infinite number of degrees of freedom, the exactness 
of the certain class of mean-field approximation in the large $N$ limit
suggests that the non-linearity is secretly inherited in quantum system with
infinite number of degrees of freedom. 
It is the non-linear mean-field equation which possesses the source of chaos 
within itself.
   
\begin{center}
{\large Acknowledgments} 
\end{center}
We are grateful to Y.Aizawa,K.Ikeda and Y.Takahashi for discussions
and comments.
One of the authors (N.T.) is supported by Research Fellowships of the Japan 
Society for the Promotion of Science for Young Scientists.

\newpage

\begin{figure}
\vspace{-1.0cm}
\centerline{
\psfig{file=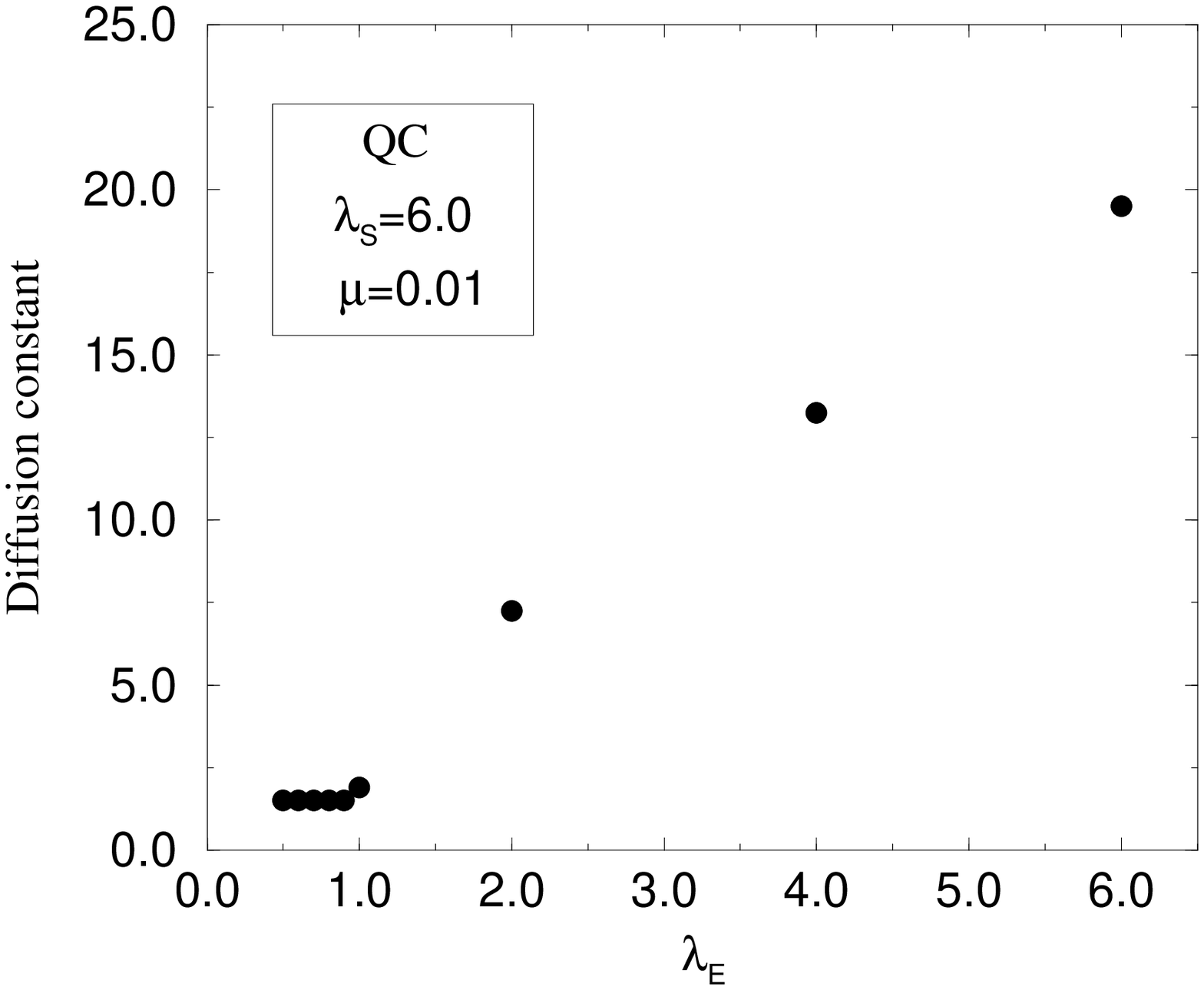,height=10cm,width=11cm}} 
\vspace{-0.5cm}
\caption
{
Diffusion constants of $S$ fixing the kick strength in the chaotic 
regime($\lambda_S=6.0$) and varying $\lambda_E$ to change the environment 
from regular to chaos.
}
\label{fig:Diffusion_Q6_Cv}
\end{figure}
\begin{figure}
\vspace{-1.0cm}
\centerline{
\psfig{file=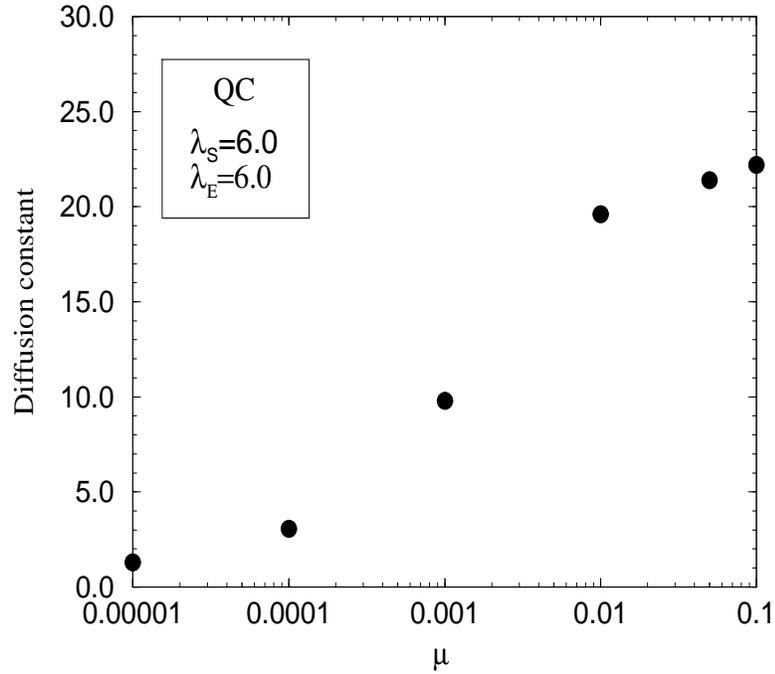,height=10cm,width=11cm}} 
\vspace{-0.5cm}
\caption
{
Diffusion constants of $S$
 fixing the kick strength of $S$ and $E$ both in the chaotic 
 regime($\lambda_S=\lambda_E=6.0$), 
and varying the coupling $\mu$.
}
\label{fig:Q6_C6_Diff.vs.mu}
\end{figure}
\begin{figure}
\vspace{-1.0cm}
\centerline{
\psfig{file=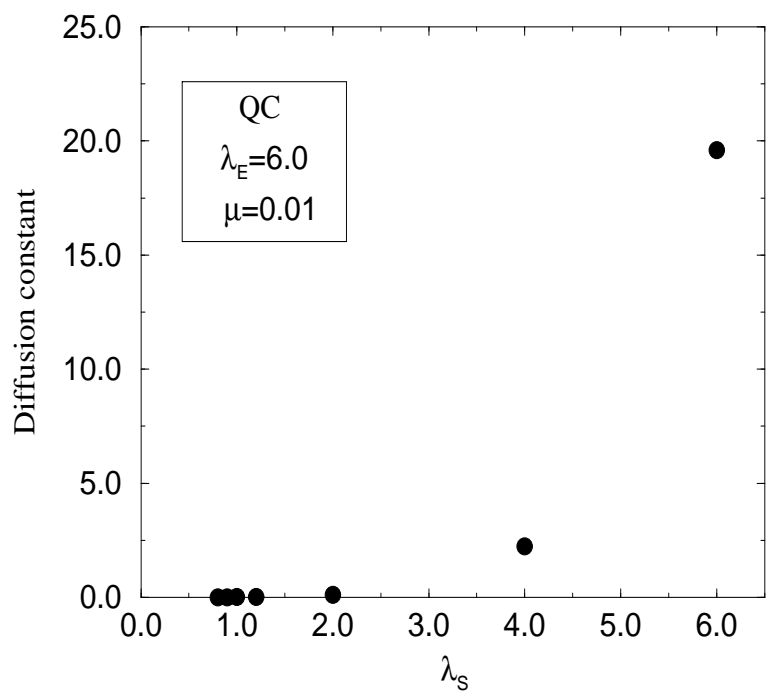,height=10cm,width=11cm}} 
\vspace{-0.5cm}
\caption
{
Diffusion constants of $S$ fixing the kick strength of $E$ in the chaotic 
 regime($\lambda_E=6.0$), and varying $\lambda_S$.
}
\label{fig:Diffusion_Qv_C6}
\end{figure}
\begin{figure}
\centerline{
\psfig{file=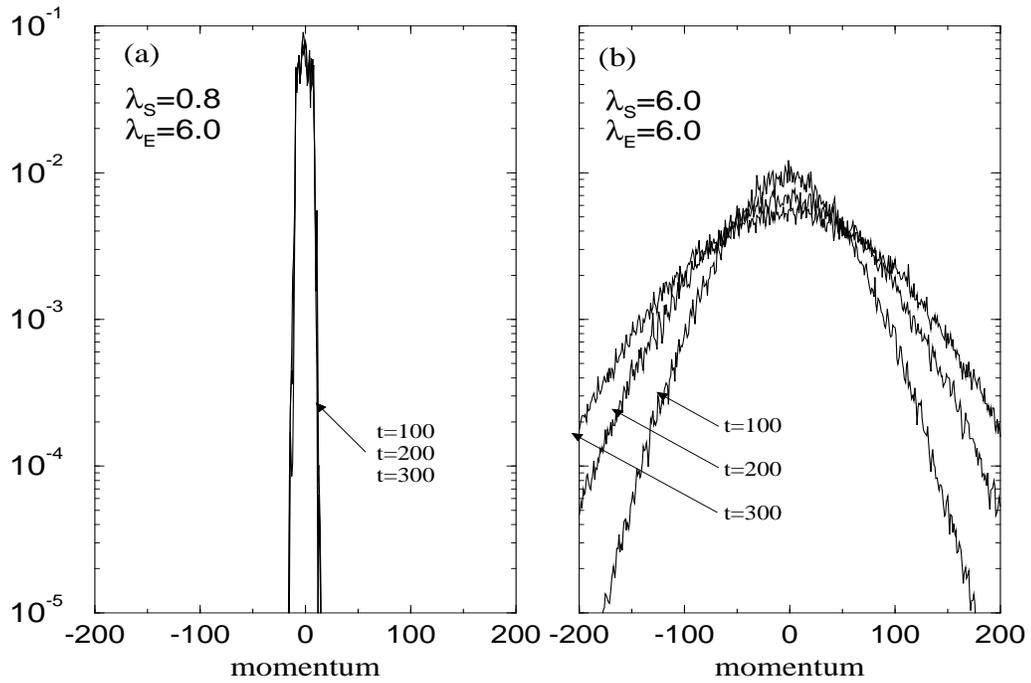,height=9cm,width=13cm}} 
\vspace{-0.0cm}
\caption
{
Time evolutions of the momentum distributions of $S$ setting the kick 
strength of $E$ in the chaotic regime($\lambda_E=6.0$) and keeping 
the system in $(a)$ the regular motion ($\lambda_S=0.8$), or $(b)$ the 
chaotic motion ($\lambda_S=6.0$).
Both of $(a)$ and $(b)$ are given by 100 ensembles average.
}
\label{fig:QC_S0.8_C6.0_moment_comp}
\end{figure}

\end{document}